\RequirePackage{fix-cm}
\documentclass[smallextended]{svjour3}       % onecolumn (second format)
\smartqed  % flush right qed marks, e.g. at end of proof
\usepackage{amssymb}
\usepackage{hyperref}
\usepackage{multirow}
\usepackage{multicol}
\usepackage{float}
\usepackage{amsmath}
\usepackage{rotating}
\usepackage{natbib}

\begin{document}

\title{Mobility-on-demand versus fixed-route transit systems: an evaluation of traveler preferences in low-income communities%\thanks{Grants or other notes
%about the article that should go on the front page should be
%placed here. General acknowledgments should be placed at the end of the article.}
}

\titlerunning{Mobility-on-demand versus fixed-route transit systems}        % if too long for running head

\author{Xiang Yan 
\and Xilei Zhao 
\and Yuan Han 
\and Pascal Van Hentenryck 
\and Tawanna Dillahunt
}

\institute{X. Yan \at
              Taubman College of Architecture and Urban Planning, University of Michigan, 2000 Bonisteel Blvd, Ann Arbor, MI 48109
%             \emph{Present address:} of F. Author  %  if needed
           \and
           X. Zhao [Corresponding Author] \at
              H. Milton Stewart School of Industrial and Systems Engineering, Georgia Institute of Technology, 755 Ferst Drive, NW, Atlanta, GA 30332\\
              Tel.: +1 443-240-0922\\
              \email{xilei.zhao@isye.gatech.edu}
          \and
          Y. Han \at
              Taubman College of Architecture and Urban Planning, University of Michigan, 2000 Bonisteel Blvd, Ann Arbor, MI 48109
%             \emph{Present address:} of F. Author  %  if needed
           \and
           P. Van Hentenryck \at
              H. Milton Stewart School of Industrial and Systems Engineering, Georgia Institute of Technology, 755 Ferst Drive, NW, Atlanta, GA 30332
          \and
          T. Dillahunt \at
              School of Information, University of Michigan, 105 S State St, Ann Arbor, MI 48109 \\
              Electrical Engineering and Computer Science, University of Michigan, 1301 Beal Avenue, Ann Arbor, MI  48109-2122
}

\date{
%Received: date / Accepted: date
}
% The correct dates will be entered by the editor

\maketitle

\begin{abstract}
Emerging transportation technologies, such as ride-hailing and autonomous vehicles, are disrupting the transportation sector and transforming public transit. Some transit observers envision future public transit to be integrated transit systems with fixed-route services running along major corridors and on-demand ridesharing services covering lower-density areas. A switch from a conventional fixed-route service model to this kind of integrated mobility-on-demand transit system, however, may elicit varied responses from local residents. This paper evaluates traveler preferences for a proposed integrated mobility-on-demand transit system versus the existing fixed-route system, with a particular focus on disadvantaged travelers. We conducted a survey in two low-resource communities in the United States, namely, Detroit and Ypsilanti, Michigan. A majority of survey respondents preferred a mobility-on-demand transit system over a fixed-route one. Based on ordered logit model outputs, we found a stronger preference for mobility-on-demand transit among males, college graduates, individuals who have never heard of or used ride-hailing before, and individuals who currently receive inferior transit services. By contrast, preferences varied little by age, income, race, or disability status. The most important benefit of a mobility-on-demand transit system perceived by the survey respondents is enhanced transit accessibility to different destinations, whereas their major concerns include the need to actively request rides, possible transit-fare increases, and potential technological failures. Addressing the concerns of female riders and accommodating the needs of less technology-proficient individuals should be major priorities for transit agencies that are considering mobility-on-demand initiatives. 

% \PACS{PACS code1 \and PACS code2 \and more}
% \subclass{MSC code1 \and MSC code2 \and more}
\end{abstract}

\section{Introduction}
\label{sec1}

When designing their transit systems, transit agencies need to carefully balance two competing goals---ridership and coverage \citep{walker2012human}. To maximize ridership under a fixed budget, an agency would avoid providing transit services to places where the demand is low; however, to ensure equitable geographic coverage and especially to accommodate the travel needs of transit-dependent populations, almost every transit operator has to make such efficiency sacrifices to some degree. More generally, the trade-off between the two competing goals results in the first-/last-mile problem of public transit, which refers to transit's inability to deliver travelers all the way from their point of origin to their destination. 

The emergence of mobility-on-demand (MOD) services, such as carsharing, bikesharing, ride-hailing (e.g., Uber and Lyft), and microtransit (e.g., Bridj, Chariot, and Via), has inspired many transit operators to incorporate these alternatives into their service suite to address these challenges, i.e., serving low-demand areas and/or providing convenient last-mile connections to transit stops. For example, many transit providers have forged partnerships with transportation network companies (mainly Lyft and Uber) to explore the potential synergy between on-demand ridesharing and transit services \citep{schwieterman,feigon2016shared}. A number of pilot projects, including eleven projects funded by the U.S. Department of Transportation's Mobility-on-Demand Sandbox Program, have been deployed by transit agencies across the United States to examine if MOD initiatives can help enhance last-mile transit connections, reduce operating costs, improve service availability, and elevate rider experiences. %Common MOD initiatives include offering subsidized Uber/Lyft rides to and from transit stations, providing on-demand rides to individuals with disabilities, replacing one or two low-ridership bus routes with subsidized Uber/Lyft ride services, and piloting microtransit services across a given service area. 

Success in these initiatives would encourage transit operators to scale up efforts to embrace MOD, opening up the possibility of designing future transit systems that fully integrate MOD with conventional public transit.\footnote{Since these initiatives are pretty recent, to our knowledge, there is not yet a comprehensive analysis of their performance. After reading some online newspaper articles and several published reports and interviewing eleven transit professionals, we found that the performance outcomes (in terms of usage and user experience) are somewhat mixed \citep{schwieterman}. Some pilots are deemed to be successful, such as the Pinellas Suncoast Transit Authority's Direct Connect program and Innisfil, Ontario, Canada's Innisfil Transit project. One example of a ``failed" experiment is the Bridj/Kansas City microtransit program, which ended after six months due to extremely low ridership. Since little research has been done to evaluate these projects, the main reasons behind the success or failure of MOD transit initiatives are still unclear.} In fact, many transit observers envision future transit systems to be integrated MOD and conventional mass transit systems (to be termed as MOD transit systems or MOD transit for the rest of the paper) featuring the synchronization of the two types of services, with large-volume mass transit (trains and buses) efficiently servicing high-demand corridors while on-demand ridesharing services covering low-density areas and filling first-/last-mile service gaps \citep{maheo2017benders,stiglic2018enhancing,yan2018integrating,shen2018integrating}.
%Compared to a conventional fixed-route system, the user experience under a MOD transit transit would differ dramatically, and transit operators need to evaluate the community support for it before switching significant proportions of its fixed-route services to MOD.

While a MOD transit system is conceptually appealing and technologically feasible, the existing knowledge regarding the traveler preferences for it is limited. When considering a switch from a predominantly fixed-route system to a MOD transit system, transit operators need to assess community support beforehand. Since public transit is often charged with equity goals of serving individuals with limited mobility options, it is especially important to pay specific attention to the travel needs and preferences of the low-income, elderly, carless, and disabled travelers. This paper aims to advance research in this area by investigating public preferences for a MOD transit system versus a fixed-route transit system, with a particular focus on disadvantaged travelers, i.e., those who are low-income, elderly, carless, and disabled. Since a truly integrated MOD transit system does not exist yet, our research approach involves conducting a stated-preference survey among residents, in two low-resource localities in Michigan--the City of Detroit and the Ypsilanti area (Ypsilanti Township and the City of Ypsilanti). 

We seek to answer three questions: Do individuals, particularly the low-income ones, prefer a MOD transit system over a fixed-route system? What factors (e.g., the socioeconomic and demographic characteristics of a respondent, the transit services they receive currently, and their use and perception of public transit and ride-hailing services) can help explain their preferences? What are the potential benefits associated with MOD transit services that individuals perceive, and what concerns do they have? Answering these questions can help transit agencies assess overall community support for MOD, identify the primary customer base for it and the benefits they receive, identify potential geographic areas for pilot deployment, and learn who might be left out by MOD transit services and the obstacles that they face.

%We conducted statistical analysis on the survey results, which is complemented with interviews with eleven transit professionals in the United States. The insights gained from these interviews facilitate the interpretation of model results and inform the discussion section of the paper.  

%Since the largest cost in providing MOD services is the driver cost involved, the idea of an integrated MOD and public transit system (to be termed as a MOD transit system in this paper) can become even more appealing as fully automated vehicles hit the road in the future. 
%the rise of MOD and the imminence of autonomous vehicle are disrupting how people travel transforming  transportation, public transit need to actively seek ways to adapt.

The remainder of this paper is organized as follows. The next section provides more background information on the concept of MOD transit systems and reviews relevant literature. Section 3 describes the data and methodology, and Section 4 presents and interprets the survey results and statistical model outputs. Section 5 applies findings to policy and discusses the limitations of this study. Section 6 concludes.

\section{The rise of mobility-on-demand, public transit, and traveler preferences}
%Since this paper primarily examines stated preference for a MOD transit system over a fixed-route system, its virtue largely rests on the assumption that Mobility-on-Demand (MOD) transit systems are realistic scenarios of future public transit systems.  
%A major motivation behind this study is the authors' 

%predict and provide?

%Public and private mobility
%Competition or complementary
%Outlook for public transit: Mobility-on-demand pilots/mobility-as-a-service concept
% Hub-and-spoke transit system vision. Autonomous vehicles
%User responses?

%declining transit ridership

%Public support for public transit and transit innovations
% The digital divide or geographical constraint
\subsection{Emerging transportation technologies and the transformation of public transit}
The emergence of mobility-on-demand services, particularly the rapid rise of ride-hailing companies such as Uber and Lyft in recent years, is disrupting the transportation sector and changing how people travel. Powered by the advances in information and communication technology, MOD serves individuals' travel needs on an as-needed basis in real time and thus provides consumers with convenience, flexibility, and cost-savings. Some transit proponents are concerned that privately-operated MOD services would siphon off transit users who have better capability to pay, thus suffocating the already struggling transit sector \citep{clewlow2017ridehailing}. On the other hand, it is widely suggested that MOD can be used to complement conventional public transit by addressing the last-mile problem and filling the transit service gaps in low-ridership areas \citep{shaheen2016lastmile}. Recognizing this potential, transit agencies have developed a variety of pilot projects, mostly through partnerships with private companies, to experiment with the idea of MOD transit services. Dozens of transit operators in the U.S. have forged partnerships with ride-haling (Uber/Lyft), microtransit (Via/Bridj), and local taxi companies to provide subsidized rides to transit stops, to substitute/complement the demand-responsive paratransit services for seniors and disabled residents, to replace one or several lower-ridership fixed-route services, and to expand services to places where transit was inadequate or nonexistent.

%Parallel to this more incremental approach of testing MOD initiatives in the form of small-scale pilot projects is the increasingly popularity of the Mobility-as-a-Service (MaaS) concept, which

%Parallel to the growth of private MOD options is the increasing popularity of the Mobility-as-a-Service (MaaS) concept, which features the integration of a variety of shared modes into a single digital platform that enables users

Parallel to the pilot deployment of small-scale MOD initiatives is the the increasingly popularity of the Mobility-as-a-Service (MaaS) concept, which visions a full integration of various mobility options into a single digital platform that enables users to make customized and multimodal travel decisions. MaaS would be made feasible by the use of information and communication technologies and the coordination of different travel modes provided by various mobility-service providers, both private and public. As the MaaS phenomenon gains momentum across the globe (primarily in Europe at present), it is likely to greatly transform the role of public transit, both in terms of its operating model and the type of services it provides to travelers \citep{hensher2017future, mulley2018workshop}. At one extreme, traditional fixed-route transit services may be replaced by privately-operated MOD services in all places except the highest-demand corridors; at the other, public transit agencies become the single MaaS provider that operates a variety of shared modes, ranging from fixed-route transit (rail and bus) services, to MOD shuttle services, and to carsharing, bikesharing, and scooter-sharing. 

%[equity requirements established under the Americans with Disabilities Act mandate transit operators to ]
%[integrate with ADA paratransit -- Boston Metropolitan Transportation Agency Example]

%lack of study on user preferences in general. Existing studies are mostly on identifying the soceconomic and demographic profiles of users, whereas reasons for not adopting shared mobility services among non-users are rarely studied \cite{shaheen2017travel}. Exceptions include some perceptions on bikesharing, etc.
%Another transformative technology is the fully connected and automated vehicles. Automation technologies are predicted to significantly reduce the money and time costs of travel, 

The future of public transit is thus full of uncertainties in the realm of market-driven technological changes (e.g., private MOD services, the MaaS concept, and the imminent autonomous vehicles) and institutional decisions made along the way as these technologies gains wider popularity. Since MOD services are expected to grow in significance, public transit agencies should actively seek opportunities to engage with them in order to keep transit being attractive to a wider population. While envisioning the future for their transit systems, transit operators need not only to seek creative approaches to improve operation efficiency and adjust its service models, but also to carefully evaluate the preferences among their constituents as to what kinds of services changes would be better received, particularly by individuals with limited mobility options \citep{sochor2015users, durand2018mobility}.

\subsection{Envisioning the future mobility-on-demand transit systems}
Given the efficiency of conventional buses and trains in moving large volumes of travelers along busy traffic corridors and the flexibility of small-sized MOD modes in serving low-demand areas, an ideal transit system, as envisioned by some transit proponents, would feature the integration and coordination of the two types of services; that is, they would complement each other by servicing distinct geographic areas instead of competing for the same customers \citep{maheo2017benders}. This integrated transit system is a natural extension of previous efforts to facilitate service integration among different transit service components, such as the connection between rail and bus lines and between the line-based transit services and demand-response services \citep{errico2013survey,wang2014DRT}. The deployment of fully connected and autonomous vehicles would further augment this vision, as automation technologies are predicted to significantly reduce the operation costs of transit (mostly driver cost) and facilitate communication and coordination among the service fleets \citep{buehler2018av}. 

In recent years, some researchers have started to explore the potential of this type of MOD transit to augment the transit network. For example, \cite{atasoy2015concept} proposed a Flexible Mobility on Demand concept that integrates taxi, shared-taxi, and mini-bus, and their simulation results demonstrated increased consumer surplus and operator profits. \cite{stiglic2018enhancing} applied a complex-system modeling approach to evaluate the benefits of integrating ride-sharing and public transit in a hypothetical metropolitan region and found that this integration can potentially enhance personal mobility and increase transit use. \cite{yan2018integrating} conducted a stated-preference survey to evaluate the traveler responses to an integrated transit system with ridesourcing and public transit, and found that such a system can help transit agencies improve level-of-service while reducing operating cost. \cite{shen2018integrating} further explored the synergy between autonomous vehicles and the public transit in Singapore, and their agent-based simulation results suggested that integrating autonomous vehicles with mass transit has the potential of enhancing service quality, reducing road use, and improving operating efficiency. These studies have demonstrated the benefits of MOD transit systems from an operations perspective, that is, MOD transit can result in higher operation efficiency and better overall service quality; however, there is a dearth of studies that focus on the travelers, particularly regarding how they perceive and would use MOD transit services.
%\citep{chen2017connecting} explored a system designs issue, the relative spatial position, of an integrated ride-hailing/fixed-route transit system.

To anticipate possible traveler responses to MOD transit, it is useful to examine empirical evidence on private MOD services, especially ride-hailing services. Several empirical studies have shown that early adopters of various types of MOD services (bikesharing, carsharing, and ride-hailing) are similar, who are often the young to middle-aged, college-educated, and moderate- to middle-income individuals living in urban areas \citep{clewlow2017ridehailing,dias2017behavioral, alemi2018uber, feigon2016shared}. It is thus natural for these individuals to become early adopters of and supporters for MOD transit services. The reasons for low usage rates of ride-hailing or MOD transit services among other population groups are nonetheless unclear and could include a combination of affordability concerns, the geographic availability or convenience of the services, lack of access (e.g., banking account, smartphone, and technology literacy) \citep{dillahunt2018getting}, and the differences in modal preferences and travel demands.

From a user's perspective, switching from a conventional fixed-route system to an integrated MOD and mass transit system can bring both benefits and costs. The most tangible benefit is the enhanced ``last-mile" access to transit services (less walking to transit stops), which has long been regarded as a major deterrent to transit use. In addition, by matching travel demand and service vehicles in real time, a MOD transit system may reduce wait time (by assigning the closest vehicle to pick up a customer) and the total travel time (by generating customized routes based on trip origin and destination) of a trip. On the other hand, in order to improve operating efficiency, a MOD transit system has to incorporate some features that may significantly deter transit, such as requiring customers to accept additional pickups and transfers (e.g., between on-demand vehicles and buses). Moreover, MOD's heavier reliance on modern technology is likely to elicit varied responses from transit riders. For tech-savvy individuals, the convenience created by technology may encourage them to use transit more, but this technology-literacy requirement can become an obstacle to MOD transit adoption for those who are less proficient with technology \citep{dillahunt2018getting, kodransky2014connecting, shaheen2017equity}.

\subsection{Mobility-on-demand transit and the disadvantaged travelers}
%Provided that public transit is often charged with equity goals, i.e., providing reliable services to disadvantaged travelers, 

%There are both tangible and intangible factors that drive individuals' preference and use of a travel alternative. "Tangible" service attributes refer to those that are functional and objectively quantifiable, whereas attributes that are abstract, subject, and more difficult to measure and quantify are termed "intangible." 

%Some of the barriers faced by low-income individuals with respect to vehicle sharing, such as lower levels of access to creidt, the Internet, and smartphones, would similarly apply to TNCs.
It is especially important to examine how the more transit-dependent disadvantaged travelers, those who are low-income, elderly, carless, and disabled, would react to a MOD transit system. Since these individuals are also more likely to lack access to technological capacity, a bank account, a smartphone, and a data plan
\citep{Pew2018}, switching from conventional public transit to MOD transit may leave these disadvantaged travelers further behind. Such equity concerns are commonly raised in discussions of MOD transit initiatives, and this issue would be particularly controversial if MOD services is proposed to replace existing fixed-route transit services. In addition, some are concerned that the requests of some minority travelers, particularly blacks, may be rejected by some drivers due to racial prejudice or unwillingness to go to certain neighborhoods \citep{ge2016racial}. 

Transit agencies have sought to find solutions to address these challenges. For example, when establishing partnerships with Uber and Lyft, transit operators often include a third-party mobility provider to separately serve wheelchair customers. When the City of Arlington, Texas contracted with Via to replace its fixed-route bus line, Via implemented a series of measures to accommodate disadvantaged travelers, such as adding prepaid debt cards as the payment option, allowing people to call in to request a ride, and including wheelchair-accessible vehicles into its service fleet. Other commonly raised solutions to access barriers for MOD services include building neighborhood access kiosks, providing data-plan subsidies for low-income travelers, and adding voice-activated mobility app features \citep{shaheen2017equity}. Moreover, several MOD transit pilot projects have reported that these equity concerns may not be as acute as many perceive. For example, the Go Centennial pilot in Colorado found that no user was denied by the program because of barriers to access.\footnote{See the technical report at \url{https://www.centennialco.gov/uploads/files/Government/Iteam/Go\%20Centennial\%20Final\%20Report_for\%20web.pdf}} A survey conducted in Arlington, Virginia found that seniors are readily willing to learn to use app-based MOD services. These findings should nonetheless be interpreted with caution, as they reflect the behavior and preferences of the MOD transit service users instead of nonusers who were excluded from these services in the first place.

%While equity concerns are commonly raised in discussions of emerging transportation technologies and their implications on public transit, knowledge regarding how these travelers perceive and would react to a MOD transit system is very limited.

%MOD transit may leave people who are already transportation-disadvantaged further behind, if it more difficult to access to MOD services for these individuals and the fixed-route services they depended on were taken away.

%Considering MOD's high reliance on technology, a common concern raised for MOD transit services is the access barrier created for individuals who are less technology-savvy or lacks access to smartphones, banking accounts, and data plans. Since these individuals also tend to be the lower-income and elderly population who are more transit-dependent, this equity concern is particularly relevant. 

On the other hand, switching from a conventional fixed-route system to a MOD transit system may also bring many benefits to transit-dependent individuals. Notably, since MOD transit has the potential to extend transit coverage areas, expand service hours, and enhance last-mile transit connectivity, it may provide an affordable option for disadvantaged travelers to get to key destinations of interest that had not been adequately served by transit before \citep{dillahunt2017uncovering}. For elderly and disabled travelers, the on-demand, real time feature of MOD transit may reduce the wait time of a trip and also give them more flexibility in scheduling their daily activities (the existing demand-responsive paratransit services usually require advanced booking). Given that MOD transit has both pros and cons compared to a fixed-route system, it is unclear whether disadvantaged travelers prefer a MOD transit system or a fixed-route one.

%it is imperative for transit agencies to accommodate the needs of these individuals when considering switching from a fixed-route transit system to a MOD one. 

\section{Data and methodology}
This paper examines traveler preferences for an integrated MOD transit over a fixed-route system, with a particular focus on disadvantaged individuals. We conducted a web-based survey hosted in Qualtrics in the City of Detroit and the Ypsilanti area, Michigan (i.e., the City of Ypsilanti and Ypsilanti Township). According to the QuickFacts data from the U.S. Census Bureau,\footnote{https://www.census.gov/quickfacts/fact/table/US/PST045218.} the City of Detroit has an estimated population of 673,104 in 2017 and the City of Ypsilanti and Ypsilanti Township combined has an estimated population of 76,388 in 2017. Both localities are low-source communities in the region with a significant proportion of the population living under poverty.\footnote{The survey was distributed to the northern part of the Ypsilanti Township, which is an area with more similar population and housing characteristics to the City of Ypsilanti than the southern part of the township. Therefore, we only discuss the statistics on the City of Ypsilanti when referring to poverty status and educational attainment associated with the Ypsilanti sample.} According to the American Community Survey 2013-2017 5-year estimates, the median household income in the city of Detroit and the city of Ypsilanti was \$27,838 and \$35,896 respectively, and the poverty rate was 37.9\% and 30.9\% respective. However, it should be noted that many of the Ypsilanti residents counted as "poor population" were actually unemployed college students who are attending the nearby University of Michigan, Ann Arbor or Eastern Michigan University. The proportion of people age 25 years and above who received a bachelor's degree or higher was only 14.2\% in Detroit, compared to a high value of 42.7\% in Ypsilanti. 

Participants were recruited from July to November 2018. We advertised the survey through a variety of means, including postal mails, flyers, posting on Nextdoor (an online forum for separate neighborhoods) and neighborhood association's email newsletters. In Detroit, we also complemented these distribution methods with in-person on-site recruitment at several public libraries and non-profit organization buildings to account for individuals who were potentially uncomfortable with digital devices or who did not have access to such devices. These efforts, which led to 170 valid survey responses, allowed us to recruit some individuals who are in the lowest-income bracket and are completely technology illiterate, and so we consider the quality of the Detroit sample data to be superior than that of the Ypsilanti sample data. A \$10 cash incentive was offered for participants recruited in person and a \$5 electronic gift card incentive was offered for other participants. We obtained a total of 497 and 534 completed responses from Ypsilanti and Detroit, respectively. After removing invalid responses, 457 and 443 responses were retained for further analysis. 

The survey solicited the following information from the respondents: their use and perception of local public transit and ride-hailing services (Uber/Lyft), their demographic and socioeconomic characteristics and home address, and their preference for a proposed MOD transit system (which was named RITMO) versus the current fixed-route system. In addition, respondents were asked if any of the following potential constraints to adopting MOD services applies to them: not owning a cellphone, owning a cellphone but not a smartphone, owning a smartphone but not a mobile data plan, not having a bank account, not having internet access at home, and having a disability that requires the accommodation of specialized vehicles. The lack of access to smartphones, internet access, and bank accounts for many and the need for wheelchair accessibility for the disabled travelers are frequently mentioned in the media and academic publications as major barriers to adopting MOD services. We also asked about cellphone ownership because we mentioned to respondents that calling in would be one of the options to request for on-demand rides (two other options are through a smartphone app and an internet web page), which was the case in the Via/Arlington, Texas partnership. Finally, the survey asked individuals to select among a list of potential benefits and drawbacks associated with the RITMO system and which ones mattered to them.

\begin{figure}[ht!]
    \centering
    \includegraphics[height=3in]{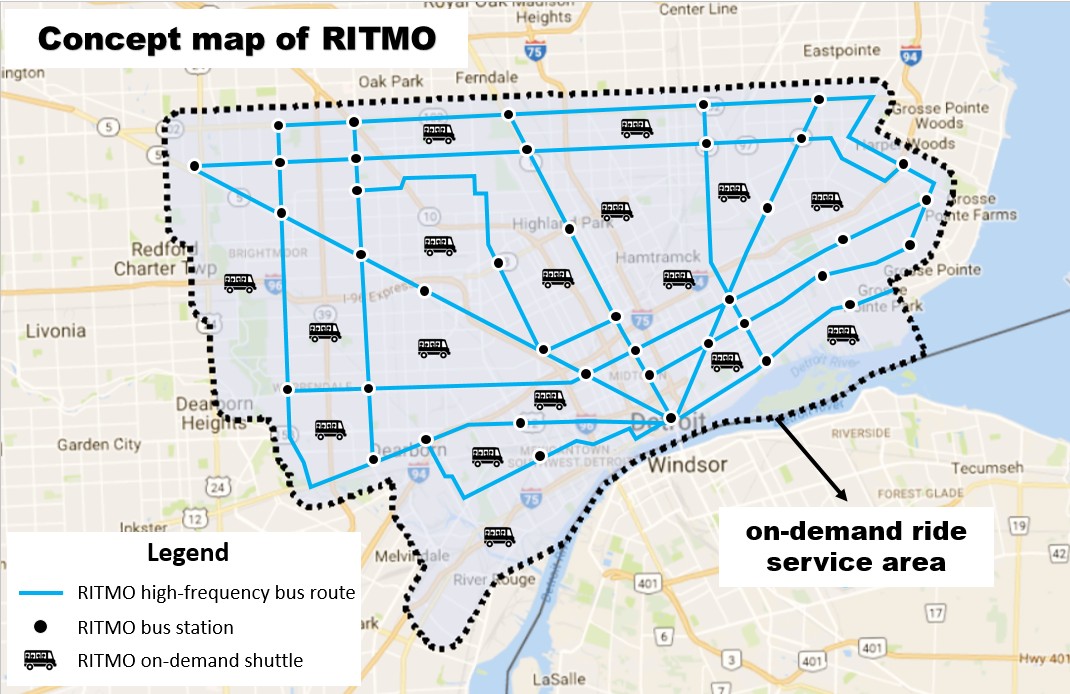}
    \caption{A concept map of the RITMO system (for Detroit respondents)}
    \label{fig:Map}
\end{figure}

The MOD transit system was introduced as following: ``The research team is testing the idea of offering a new type of transit service, called RITMO, in Detroit. This service features rapid and high-frequency RITMO buses running in major corridors and on-demand RITMO shuttles serving the outer area. Below is a concept map of RITMO (see Fig. 1)." The survey then further described RITMO with both text and image illustrations, highlighting that the RITMO on-demand shuttles can be requested by using a smartphone app, or the internet, or with a phone call. Also, respondents were told that the RITMO shuttle services will not be door-to-door, and instead they would be picked up or dropped off at a street corner close to their points of origin and destination; RITMO shuttles would also pick up passengers who shared similar origins and destinations along the way (see Fig. 2). To illustrate the travel-experience changes under RITMO versus the current fixed-route system, respondents were then presented with two sets of before-after image comparisons between a bus trip and a hypothetical RITMO trip. One set describes a short trip that would involve an on-demand shuttle only and the other describes a longer trip with a transfer between an on-demand shuttle and a bus. The survey followed with the main question of this study: ``Compared to a fixed-route bus system, do you prefer the RITMO system?" 

\begin{figure}[ht!]
    \centering
    \includegraphics[height=3.5in]{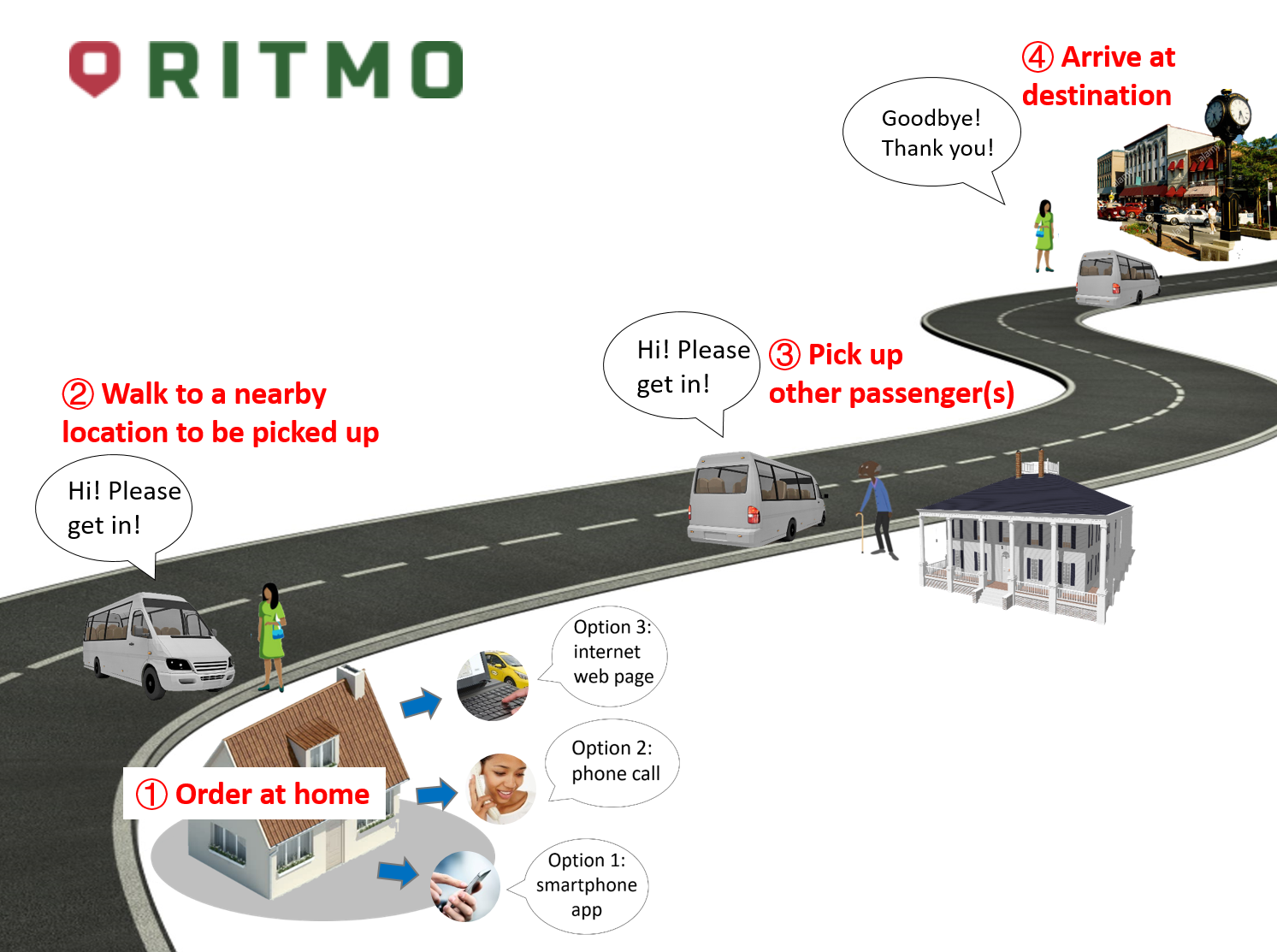}
    \caption{An illustration of riding with RITMO shuttles}
    \label{fig:Illustration}
\end{figure}

We constructed a five-level Likert-scale response variable based on answers to this question (``1" is ``I strongly prefer a fixed-route system over RITMO", ``2" is ``I sort of prefer a fixed-route system over RITMO", ``3" is ``I am not so sure", ``4" is ``I sort of prefer RITMO over a fixed-route system", and ``5" is ``I strongly prefer RITMO over a fixed-route system"). Therefore, larger values represent stronger preference for MOD over fixed-route. Since the response variable is measured using an ordinal scale, the ordered logit model was applied to analyze it. We hypothesized respondents' preference toward RITMO versus fixed-route to be shaped by a host of variables, which includes their demographic and socioeconomic characteristics, the current bus services available at their residence, and their current use and perception of the bus services and ride-hailing services. In particular, we were interested in examining if the disadvantaged travelers (those who were low-income, disabled, carless, and elderly), and those who may face access barriers to adopting MOD (e.g., do not have a smartphone and/or a mobile data plan) would have a stronger preference for fixed-route than other populations. 
%An ordered logit model ... [or generalized ordered logit model]

%Instead of evaluating traveler responses to different service attributes of a travel option, we focus on examining travelers' overall perception toward MOD transit versus traditional fixed-route services. In this survey, we constructed five-level Likert-scale response question that asks respondents if they prefer a proposed MOD transit system over the existing fixed-route system. ... 

%Our ordered logit model thus takes the following functional form:
Most variables of interest, except the bus services available to each respondent, were directly obtained from the survey data. We computed three spatial indicators to measure the transit services that respondents received from their home address, including a dummy variable that indicates if the closest bus stop to a respondent's home is within one quarter mile, the number of buses passing by within a quarter mile buffer of a respondent's home during the morning peak hours, and the transit accessibility to jobs. To compute the first two measures, we geocoded the home address of respondents using the Google Maps Geocoding API, obtained bus scheduling information from Google Transit Feeding Services (GTFS), and performed simple geoprocessing functions such as Near, Buffer, and Dissolve in ESRI's ArcMap. For the last measure, transit accessibility to jobs, we calculated the number of jobs reachable within 45 minutes of transit travel time based on the 2015 Longitudinal Employer-Household Dynamics Origin-Destination Employment Statistics data and the GTFS file (during the morning peak hours).

%[Differ from a discrete choice modeling study in that service attributes regarding MOD transit may be difficult to quantify; measure overall modal preferences

\section{Data analysis and results}
\subsection{Descriptive statistics}
Table 1 presents the descriptive profile of the two samples. Overall, we found that survey respondents in both localities were strongly in favor of the RITMO transit system (i.e., an integrated system of fixed-route bus and on-demand ridesharing services). A majority of respondents (nearly 60 percent in Detroit and over 70 percent in Ypsilanti) strongly or sort of preferred the RITMO system over the existing fixed-route transit system, and the opposite only applied to a small fraction of them (about 15 percent in Detroit and a little over 10 percent in Ypsilanti). Compared to Ypsilanti respondents, a higher proportion of the Detroit respondents suggested that they were ``not so sure" between the two systems, which may result from a higher level of uncertainty toward modern technology and distrust of public entities \citep{kodransky2014connecting}. 

% Please add the following required packages to your document preamble:
% \usepackage{multirow}
\begin{table}[ht!]
\caption{Sample characteristics ($N=443$ for Detroit sample and $N=457$ for Ypsilanti sample)}
\footnotesize
\resizebox{1\textwidth}{!}{% <------ do not forget this %
\begin{tabular}{llcccc}
\hline
\multirow{2}{*}{Variable}           & \multirow{2}{*}{Categories}             & Frequency       & Proportion       & Frequency        & Proportion        \\ \cline{3-6} 
                                    &                                         & \multicolumn{2}{c}{Detroit sample} & \multicolumn{2}{c}{Ypsilanti sample} \\ \hline
Preference for RITMO transit versus & Strongly prefer RITMO over fixed-route  & 137             & 30.93\%          & 171              & 37.42\%           \\
fixed-route             & Sort of prefer RITMO over fixed-route   & 128             & 28.89\%          & 161              & 35.23\%           \\
                                    & Not so sure                             & 114             & 25.73\%          & 74               & 16.19\%           \\
                                    & Sort of prefer fixed-route over RITMO   & 47              & 10.61\%          & 40               & 8.75\%            \\
                                    & Strongly prefer  fixed-route over RITMO & 17              & 3.84\%           & 11               & 2.41\%            \\
Gender                              & Male                                    & 249             & 56.21\%          & 240              & 52.63\%           \\
                                    & Female                                  & 194             & 43.79\%          & 216              & 47.37\%           \\
Age                                 & Under 25                                & 32              & 7.22\%           & 53               & 11.62\%           \\
                                    & 25 - 29                                 & 100             & 22.57\%          & 82               & 17.98\%           \\
                                    & 30 - 39                                 & 138             & 31.15\%          & 153              & 33.55\%           \\
                                    & 40 - 49                                 & 75              & 15.93\%          & 72               & 15.79\%           \\
                                    & 50 - 59                                 & 47              & 10.61\%          & 73               & 16.01\%           \\
                                    & 60 - 69                                 & 44              & 9.93\%           & 18               & 3.95\%            \\
                                    & 70 or over                              & 7               & 1.58\%           & 5                & 1.10\%            \\
Black                               & Yes                                     & 236             & 53.27\%          & 38               & 8.33\%            \\
                                    & No                                      & 207             & 46.73\%          & 419              & 91.89\%           \\
Household income                    & Less than \$25,000                      & 184             & 41.53\%          & 49               & 10.75\%           \\
                                    & $25,000 - $49,999                       & 104             & 23.48\%          & 126              & 27.63\%           \\
                                    & $50,000 - $74,999                       & 64              & 14.45\%          & 80               & 17.54\%           \\
                                    & $75,000 - $99,999                       & 37              & 8.35\%           & 90               & 19.74\%           \\
                                    & $100,000 - $124,999                     & 21              & 4.74\%           & 70               & 15.35\%           \\
                                    & $125,000 - $149,999                     & 18              & 4.06\%           & 30               & 6.58\%            \\
                                    & \$150,000 or more                       & 15              & 3.39\%           & 11               & 2.41\%            \\
Education attainment                & Less than high school                   & 31              & 7.31\%           & 2                & 0.45\%            \\
                                    & High school graduate                    & 201             & 47.41\%          & 40               & 8.91\%            \\
                                    & Professional degree                     & 63              & 14.86\%          & 124              & 27.62\%           \\
                                    & Bachelor degree                         & 85              & 20.05\%          & 249              & 55.46\%           \\
                                    & Master's/Doctorate                      & 44              & 10.38\%          & 34               & 7.57\%            \\
Car ownership                       & Yes                                     & 301             & 67.95\%          & 401              & 87.75\%           \\
                                    & No                                      & 142             & 32.05\%          & 56               & 12.25\%           \\
Primary travel mode                 & Public bus                              & 117             & 26.41\%          & 156              & 34.14\%           \\
                                    & Other                                   & 326             & 73.59\%          & 301              & 65.86\%           \\
Ride-hailing knowledge and use      & Never heard of or used Uber/Lyft        & 223             & 50.34\%          & 130              & 28.45\%           \\
                                    & Used Uber/Lyft at least once            & 220             & 49.66\%          & 327              & 71.55\%           \\
Have a cellphone                    & Yes                                     & 408             & 92.10\%          & 446              & 97.59\%           \\
                                    & No                                      & 35              & 7.90\%           & 11               & 2.41\%            \\
Have a smartphone                   & Yes                                     & 385             & 86.91\%          & 410              & 89.72\%           \\
                                    & No                                      & 58              & 13.09\%          & 47               & 10.28\%           \\
Have a mobile data plan             & Yes                                     & 394             & 88.94\%          & 393              & 86.00\%           \\
                                    & No                                      & 49              & 11.06\%          & 64               & 14.00\%           \\
Have Internet access at home        & Yes                                     & 368             & 83.07\%          & 438              & 95.84\%           \\
                                    & No                                      & 75              & 16.93\%          & 19               & 4.16\%            \\
Have a bank account                 & Yes                                     & 371             & 83.75\%          & 451              & 98.69\%           \\
                                    & No                                      & 72              & 16.25\%          & 6                & 1.31\%            \\
Have a disability                   & Yes                                     & 40              & 9.03\%           & 4                & 0.88\%           \\
					& No                                     & 403             & 90.97\%          & 453              & 99.12\%           \\ \hline
\end{tabular}

}
\end{table}

Our survey samples were not representative of the populations from which they were drawn. By comparing our samples with the Census Data (i.e., American Community Survey 2013-2017 5-year estimates), we found that males, non-black population, young adults (age between 25 to 40), college graduates, and transit riders were overly represented in both samples.\footnote{Census Data Statistics for Detroit: 46.3\% adults are males, 79.1\% of the population are black, 34.5\% adults are between the age of 25 to 44, 14.3\% of the population 25 years and older have a bachelor degree or above, 8.2\% of workers use public transit for commuting. Census Data Statistics for Ypsilanti (the City of Ypsilanti and Ypsilanti Township combined): 48.8\% adults are males, 30.6\% of the population are black, 26.1\% adults are between the age of 25 to 44, 33.9\% of the population 25 years and older have a bachelor degree or above, 4.6\% of workers use public transit to commute.} In addition, the Ypsilanti sample contained too few responses from the low-income (household income below \$25,000), black, aged (over 60 years old), and disabled people relative to their population shares. Over 50 percent of the Detroit respondents answered that they either have not heard of Uber/Lyft or never used them, whereas less than 30 percent of the Ypsilanti respondents stated the same. These percentages are much lower than those reported in several recent national-scale surveys on ride-hailing (e.g., \citet{clewlow2017ridehailing,smith2016shared}), which found that about 75\% of the U.S. population had not used Uber/Lyft before in 2015. Possible sources of discrepancy include sample bias and the rapid market penetration of ride-hailing services in recent years (the other surveys were conducted in 2015 and 2016 whereas our survey was conducted in 2018). 

%The Detroit sample contains fair amounts of individuals who may face access challenges to adopt on-demand ride services, i.e., limited or no access to bank account, smartphone, data plan, or who have disabilities. By contrast, the Ypsilanti sample only contains six responses without a bank account and four disabled individuals, which would prevent us from making meaningful statistical inferences for these population groups. Again, 
It appeared that the percentage of individuals facing potential access barriers to adopt mobility-on-demand is somewhat lower in our sample than the general U.S. population. For example, about 13 percent of the Detroit respondents and 10 percent of the Ypsilanti respondents reported that they did not have a smartphone respectively. However, surveys conducted by the Pew Research Center from 2008 to 2018 suggested that about 33\% of U.S. adults do not own a smartphone as of November 2016 \citep{Pew2018}. These results suggest that the overall support for MOD transit is likely to have a positive bias in our sample; that is to say, we expect that a lower percentage of the population would prefer MOD transit over fixed-route and a higher percentage of the population would prefer the opposite.

Furthermore, though not presented here, results from some cross-tabulation analysis showed that individuals with MOD access challenges are more likely to be the elderly and the low income, and those who are less-educated, which is consistent with the findings from the Pew Research Center surveys. Together with the fact that these individuals tend to use transit more frequently than other population groups (e.g., More than 50 percent of the Detroit respondents with a household income of less than \$25,000 used bus five times or more in the pat week, whereas only about 15 percent of respondents with higher income levels did the same), these findings suggest that addressing the access barriers to MOD for disadvantaged travelers should be a priority for the development of MOD transit systems.

Table 2 presents the descriptive statistics for the three spatial indicators that measure the level of transit service received by each respondent. These results show that large variations exist across individuals in terms of the current transit services they receive and that the overall transit service received by Detroit respondents are higher than that by Ypsilanti respondents. 

\begin{table}[ht!]
\caption{Descriptive statistics for the transit-service-related indicators}
\centering
\footnotesize
\resizebox{1\textwidth}{!}{% <------ do not forget this %
\begin{tabular}{p{3.5cm}|p{2cm}|c|c|c|c}
\hline
\multirow{2}{*}{Variable}                                            & \multirow{2}{*}{Variable type} & \multicolumn{2}{c|}{Detroit sample} & \multicolumn{2}{c}{Ypsilanti sample} \\ \cline{3-6} 
                                                                     &                                & Mean        & SD    & Mean         & SD     \\ \hline
Live within a quarter mile to a bus stop                             & Dummy (No=0, Yes=1)            & 0.41        & 0.49                  & 0.34         & 0.48                   \\ \hline
No. of buses passing by within a quarter mile of a respondent's home & Continous                      & 12.21       & 24.02                 & 2.28         & 3.40                   \\ \hline
No. of jobs reachable within 45 min of transit travel time       & Continous                      & 17685.44    & 25718.60              & 2743.82      & 1470.33                \\ \hline
\end{tabular}
}
\end{table}

%females are underrepresented in both samples, both of them were made up of fairly diverse population groups and thus can be used to examine the preference heterogeneity across population segments. In particular, we have collected a decent number of responses from individuals across different age and income groups, 

%Federal Deposit Insurance Corporation estimated 8 percent of U.S. households are unbanked (FDIC, 2014).
%64\% OF Americans own smartphones (Pew Research Center, 2015); smartphone access varies more by age than by income, 27\% of adults over aged 65 have a smartphone; ownership among African American teens is higher than that among either white or Hispanic teens..

%As of November if 2016, 77\% of U.S. adults owned a smartphone, up from 35\% in 2011 (Street et al., 2017).However, there are still discrepancies in use across demographics with use lowest among those 65 years and older (42 percent), without a high school diploma (54 percent), with incomes below \$30,000 (64 percent), and those living in rural areas (67 percent) (Street et al., 2017).

%About 10\% of U.S. population has a physical limitation of some kind that requires the use of accessible vehicles (U.S. Census Bureau, 2012).

\subsection{Ordered logit model specification}
We further applied ordered logit models to examine the determinants of individuals' preference for MOD transit versus fixed-route. We hypothesized individual preferences to be shaped by a battery of factors, including the current fixed-route bus services available at their home, their use and perception of public transit and ride-hailing services, their demographic and socioeconomic characteristics, and any access constraints they face to adopting mobility-on-demand. In particular, we examined if disadvantaged travelers and less tech-savvy individuals have a stronger preference for fixed-route transit than other population groups. Thus the functional form of the final model (descriptions of the independent variables are shown in Table 3) is:
\begin{multline}
    y =  \text{Male} + \text{AgeBelow40} + \text{AgeAbove60} + \text{LowInc} + \text{Black} + \text{College} \\
    + \text{Carless}
    + \text{BusTravel} + \text{NoRidehail} + \text{NoBank} + \text{NoSmaPhone} \\ + \text{NoDataPlan} + \text{NoBank}
    + \text{NoCellPhone} + \text{NoInternet} \\
    + \text{Disability}
    + \text{BusStopQrtMi} + \text{BusJobAcc}  + \text{BusFreq} + e.
\end{multline}

We fit three ordered logit models in Stata with the above specification--one for the Detroit data, one for the Ypsilanti data, and one for the pooled data. In the Ypsilanti model, NoBank, NoCellPhone, NoInternet, and Disability were omitted due to the small sample size associated with these variables, and BusFreq was omitted because of its high correlation with BusStopQrtMi (correlation index is 0.85). Detroit, a dummy variable that takes a value of one if the respondent belongs to the Detroit sample, was added into the pooled data model. 

\begin{table}[ht!]
\caption{Description of independent variables}
\footnotesize
\resizebox{1\textwidth}{!}{% <------ do not forget this %
\begin{tabular}{l|l|p{8.5cm}}
\hline
Variable code & Type      & Value and description                                                                         \\ \hline
Male          & Dummy     & 1=Male; 0=Female                                                                             \\ \hline
AgeBelow40    & Dummy     & 1=Age is below 40; 0=Age is 40 or above                                                       \\ \hline
AgeAbove60    & Dummy     & 1=Age is 60 or above; 0=Age is below 60                                                       \\ \hline
LowInc        & Dummy     & 1=Household income is below \$25,000; 0=Household income is \$25,000 or above                   \\ \hline
Black         & Dummy     & 1=Race is black; 0=Race is non-black                                                          \\ \hline
College       & Dummy     & 1=Have a bachelor degree or above; 0=Do not have a bachelor degree                            \\ \hline
Carless       & Dummy     & 1=Do not Have a car; 0=Have a car                                                             \\ \hline
BusTravel     & Dummy     & 1=Primary travel mode is bus; 0=Primary travel mode is not bus                                \\ \hline
NoRidehail    & Dummy     & 1=Never heard or used ridehailing before; 0=Used ridehailing at least once                    \\ \hline
NoBank        & Dummy     & 1=Do not have a bank account; 0=Have a bank account                                           \\ \hline
NoSmaPhone    & Dummy     & 1=Have a cellphone but it is not a smartphone; 0=Have a smartphone                            \\ \hline
NoDataPlan    & Dummy     & 1=Have a smartphone but not a mobile data plan; 0=Have a data plan                            \\ \hline
NoCellPhone   & Dummy     & 1=Have no cellphone; 0=Have a cellphone                                                       \\ \hline
NoInternet    & Dummy     & 1=Have no internet access at home; 0=Have interenet access at home                            \\ \hline
Disability    & Dummy     & 1=Have a disability that require wheelchair accessibility; 0=Have no disability               \\ \hline
BusStopQrtMi  & Dummy     & 1=Live within a quarter mile to a bus stop; 0=Do not live within a quarter mile to a bus stop \\ \hline
BusJobAcc     & Continous & No. of jobs reachable within 45 minutes of transit travel time                                \\ \hline
BusFreq       & Continous & No. of buses passing by within a quarter mile of a respondent's home                          \\ \hline
\end{tabular}
}
\end{table}

Two additional technical details regarding the ordered logit model warrants further discussion. First, we checked the proportional odds assumption. The proportional odds assumption constraints the coefficients that describe the relationship between each pair of response categories to be the same, e.g., coefficients that describe the relationship between the lowest versus all higher categories of the response variable is the same as those that describe the relationship between the highest versus all lower categories of the response variable. The violation of this assumption often leads modellers to use a generalized ordered logit model instead. We tested the proportion odds assumption for all three models and found that only the pooled data model violated this assumption at the 0.05 level of significance. After examining the results of the generalized ordered logit model for the pooled data, we decided to stick with the ordered logit model due to two reasons: First, the generalized ordered logit model generated little additional substantive insights; Second, it resulted in some case outcomes with a predicted probability that is less than zero \citep{williams2016generalized}. 

Second, we considered whether to collapse categories of outcome with small sample sizes or not. As shown in Table 1, the sample size for the ``strongly prefer fixed-route over RITMO" response category is very small for both the Detroit sample ($N=17$) and the Ypsilanti sample ($N=11$), which motivated us to consider the option of combining these responses with the closest response category (i.e., ``sort of prefer fixed-route over RITMO"). We thus refitted the three ordered logit models with four categories of outcome, but the model outputs were very similar to those of the original models with five categories of outcome. Since some statisticians have raised concerns of obtaining biased effect estimates when response categories were collapsed \citep{stromberg1996collapsing}, we decided to keep the original five response categories.

\subsection{Ordered logit model outputs and interpretations}
Table 4 presents the model outputs for the three ordered logit models. As shown at the bottom of the table, the likelihood ratio chi-square test showed that all three models are significant improvements compared to the intercept-only null model. The McFadden pseudo r-squared values were relatively low (between 0.08 and 0.12), which is common for this type of model. Overall, the outputs of the three models were reasonably similar, suggesting that the results are robust to variations in local contexts, i.e., differences in population and socioeconomic characteristics and transit systems. In the pooled data model, the Detroit dummy variable was not statistically significant at the 0.05 level. While not shown in this paper, we also ran another pooled data model with the Detroit dummy as the only independent variable; in this model, the coefficient on the Detroit dummy was negative and significant at the 0.05 level, which suggests that there are significant preference differences between the Detroit sample and the Ypsilanti sample. The fact that the Detroit dummy variable became insignificant in the final model implies that such preference differences were captured by other independent variables, which adds further evidence suggesting that the final model was adequately specified. 

\begin{table}[ht!]
\caption{Ordered logit model outputs}
\footnotesize
\resizebox{1\textwidth}{!}{% <------ do not forget this %
\begin{tabular}{lcccccc}
\hline
\multirow{2}{*}{Variable}                & \multicolumn{2}{c}{Detroit data} & \multicolumn{2}{c}{Ypsilanti data} & \multicolumn{2}{c}{Pooled data} \\ \cline{2-7} 
                                          & Coeff.   & S.E.   & Coeff.    & S.E.    & Coeff.   & S.E.  \\ \hline
Male                                      & 0.395*        & 0.193            & 0.536**        & 0.196             & 0.352**       & 0.132           \\
AgeBelow40                                & 0.344         & 0.221            & -0.0513        & 0.21              & 0.168         & 0.149           \\
AgeAbove60                                & 0.470         & 0.347            & -1.273*        & 0.523             & -0.229        & 0.277           \\
LowInc                                    & -0.148        & 0.229            & 0.0680         & 0.351             & -0.110        & 0.189           \\
Black                                     & 0.258         & 0.202            & 0.210          & 0.344             & 0.218         & 0.168           \\
College                                   & 0.592**       & 0.224            & 1.180**        & 0.225             & 0.837**       & 0.154           \\
Carless                                   & 0.441         & 0.262            & 1.233**        & 0.356             & 0.693**       & 0.207           \\
BusTravel                                 & -0.0852       & 0.245            & -0.0508        & 0.213             & -0.0717       & 0.155           \\
NoRidehail                                & -0.677**      & 0.2              & -0.794**       & 0.232             & -0.780**      & 0.148           \\
NoSmaPhone                                & -0.483        & 0.293            & 0.170          & 0.354             & -0.125        & 0.219           \\
NoDataPlan                                & -0.900**      & 0.317            & -0.387         & 0.288             & -0.540**      & 0.204           \\
NoBank                                    & 0.170         & 0.293            &                &                   & -0.116        & 0.278           \\
NoCellPhone                               & -0.195        & 0.425            &                &                   & 0.303         & 0.359           \\
NoInternet                                & -0.197        & 0.285            &                &                   & -0.343        & 0.233           \\
Disability                                & -0.145        & 0.34             &                &                   & -0.379        & 0.313           \\
BusStopQrtMi                              & -0.572*       & 0.235            & -0.395*        & 0.196             & -0.517**      & 0.151           \\
BusJobAcc (divided by 10,000)             & -0.184**      & 0.059            & -3.98**        & 0.682             & -0.181**      & 0.059           \\
BusFreq                                   & -0.001        & 0.006            &                &                   & -0.002        & 0.006           \\
Detroit                                   &               &                  &                &                   & 0.136         & 0.168           \\
                                          &               &                  &                &                   &               &                 \\
Cutpoint 1                                & -4.462**      & 0.485            & -6.043**       & 0.601             & -4.586**      & 0.369           \\
Cutpoint 2                                & -2.797**      & 0.41             & -4.213**       & 0.516             & -2.875**      & 0.312           \\
Cutpoint 3                                & -1.108**      & 0.384            & -2.889**       & 0.492             & -1.398**      & 0.296           \\
Cutpoint 4                                & 0.291         & 0.382            & -0.966*        & 0.471             & 0.175         & 0.292           \\
                                          &               &                  &                &                   &               &                 \\
Observations                              & \multicolumn{2}{c}{415}          & \multicolumn{2}{c}{411}            & \multicolumn{2}{c}{826}         \\
Log likelihood (Null Model)               & \multicolumn{2}{c}{-592.67}      & \multicolumn{2}{c}{-546.43}        & \multicolumn{2}{c}{-1148.12}    \\
Log likelihood                            & \multicolumn{2}{c}{-546.61}      & \multicolumn{2}{c}{-482.81}        & \multicolumn{2}{c}{-1058.61}    \\
Likelihood ratio chi-square statistic     & \multicolumn{2}{c}{92.11}        & \multicolumn{2}{c}{127.24}         & \multicolumn{2}{c}{179.01}      \\
p-value                                   & \multicolumn{2}{c}{0.00}         & \multicolumn{2}{c}{0.00}           & \multicolumn{2}{c}{0.00}        \\
Pseudo R-squared                          & \multicolumn{2}{c}{0.08}         & \multicolumn{2}{c}{0.12}           & \multicolumn{2}{c}{0.08}        \\ \hline
Standard errors in parentheses            &               &                  &                &                   &               &                 \\
** p\textless{}0.01, * p\textless{}0.05 &               &                  &                &                   &               &                
\end{tabular}

}
\end{table}

The subsequent analysis examines the two sample-specific models only. We interpreted the statistical significance of the coefficient estimates and their signs. It should be noted that a positive coefficient indicates a higher probability of \textit{choosing a response category coded with a larger value}, which means a stronger preference for MOD transit or a weaker preference for fixed-route, instead of suggesting a higher probability of \textit{choosing MOD transit over fixed-route}. Independent variables that were significant at the 0.05 level in both the Detroit model and the Ypsilanti model include Male, College, NoRidehail, BusStopQrtMi, and BusJobAcc. We found that males and college graduates are more likely to select a response category of larger values than females and individuals without a bachelor degree. For example, the Detroit model showed that males were 1.47 (e\textsuperscript{0.395}) times more likely than females to hold a more preferable view on MOD transit. Based on the comments we received from the survey respondents, we speculated that the preference differences between males and females may result from the fact that females were more concerned about potential safety/privacy issues associated with ridesharing, e.g., feeling uncomfortable sharing rides with strangers in small-size on-demand vehicles. This is consistent with other studies that found a distrust of strangers and a concern for safety are major barriers to the sharing economy \citep{tussyadiah2018drivers}. College graduates may have a stronger preference for MOD transit because they were more tech savvy and also were more adaptable to new and innovative ideas.  

Individuals who had never heard of or used ride-hailing before and were better served by the current fixed-route system (i.e., living within walking distant to a transit stop and being able to reach more opportunities) were less likely to select a response category that indicates a stronger preference for MOD transit over fixed-route transit. For example, from the Ypsilanti model we observed that, compared to individuals living more than a quarter mile to the nearest bus stop, those living within a quarter mile were only 0.67 (e\textsuperscript{-0.395}) times as likely to choose a response category with a larger value. These findings were not surprising. Individuals who had no experience with ride-hailing services tend to be people who are not tech savvy, or who are reluctant to change travel habits and to experience new things, or who hold a negative perception of ride-hailing, and so it is natural for them to be less favorable towards MOD transit than other individuals. Previous studies have found that technology proficiency and perceived ease of use are major factors that impact individuals' willing to participate in the sharing economy \citep{hsiao2018role}. It is also natural for individuals who received better transit services from the existing fixed-route transit system to express less support for change (i.e., from a fixed-route to a MOD transit system). 

Some variables were significant in one of the two sample-specific models but not the other. The Detroit model showed that compared to individuals having a data plan with their smartphones, those who had a smartphone but not a data plan were less likely to indicate a stronger preference for MOD transit over fixed-route. By contrast, the variable NoDataPlan shared a negative sign but was not statistically significant at the 0.05 level in the Ypsilanti model. On the other hand, the Ypsilanti model showed that aged (60 years old or above) respondents had weaker preference for MOD transit than the middle-aged (the reference age group, between 40 and 60 years old) and that carless individuals had a stronger preference for MOD transit. By contrast, neither variables were statistically significant at the 0.05 level in the Detroit model. Also, while Carless had a positive sign in both models, AgeAbove60 had a negative sign in the Ypsilanti model but a positive sign in the Detroit model. A possible explanation for the results on AgeAbove60 is that seniors living in the Ypsilanti area are more satisfied with the current fixed-route services they receive than seniors living in the city of Detroit.

We found that the preferences of young adults (age between 18 and 40) were not significantly different from middle-aged individuals, and variable AgeBelow40 had a positive sign in the Detroit model but a negative sign in the Ypsilanti model. This finding implies the reported higher use of mobility-on-demand services among young adults may not be due to individual preferences but result from the geographic availability of these services at the places where they live. In addition, individual preferences did not vary significantly by income, race, or primary travel mode.

Furthermore, the results showed that the preferences for MOD versus fixed-route transit among individuals without a bank account (NoBank), with a disability (Disability),  without a smartphone or even a cellphone (NoCellPhone, NoSmaPhone), or without internet access at home (NoInternet) were not significantly different from other individuals. These findings are somewhat unexpected, as one would expect individuals facing these potential access barriers to adopt mobility-on-demand to have a less preferable view toward MOD transit than the rest of the population. These results may be an artifact of how we described the proposed RITMO system in the survey. The survey did not mention possible payment options and if the on-demand shuttles would be wheelchair accessible; therefore, individuals without a bank account and or a disability may simply assumed that, like the existing transit system, cash payments would be allowed and wheelchair accessibility would be provided under the RITMO system. Also, as mentioned above, we told the respondents that there would be three options to request for a on-demand ride---through a smartphone app, an internet page, or a phone call. We provided no further information on the convenience of use regarding the three options. Since very few of our respondents (13 individuals in the Detroit sample and zero in the Ypsilanti sample) had access to none of the three options, i.e., they have at least one of the following---a cellphone, a smartphone with mobile data, or internet access, they may not perceive accessing MOD services as a problem. This points to a major limitation of our study, that is, there is a lack of representation of the least technology-savvy individuals in our survey sample.  On the positive side, however, these results also suggest that allowing passengers to request rides with phone calls may effectively mitigate the technology-access problem for adopting MOD transit.
%In our description of the RITMO system, we mentioned to the respondents that on-demand rides can be requested with three options (a smartphone app, internet page, or phone call) but did not describe possible payment options and if the on-demand shuttles would be wheelchair accessible. Thus the parameter estimates for NoSmaPhone is understandable, and we provide two possible explanations for the results on NoBank and Disablility. First, the results may be subject to a hypothetical bias; that is to say, given the hypothetical nature of the proposed MOD transit systems and the lack of real experience in using it, respondents may not have recognized these access barriers as an issue. It may also be that respondents simply assumed the need for cash payments and that wheelchair accessibility would be accommodated. 

To conclude, our statistical analysis generates insights regarding the preference differences for MOD transit across population segments. First, we obtained strong evidence suggesting that people who are more likely to resist a switch from a fixed-route transit system to a MOD one include females, individuals without a college degree, individuals who had never heard of or used ride-hailing services, and individuals who are currently well served by the existing fixed-route services. We also found that individuals without a car and those without a data plan tend to hold a less favorable view on MOD transit, although the evidence is less conclusive (i.e., results are only statistically significant in one model but not the other). Moreover, there appeared to be little preference difference across age, income, and racial groups, as the models resulted in statistically insignificant coefficients or even contradictory signs. In addition, individuals who primarily rely on public transit for travel shared a similar preference pattern with people traveling with other models. Finally, somewhat surprisingly, we found that individuals without a bank account, without a smart phone, with internet access at home, or with a disability are no more or less likely to hold a stronger preference for MOD transit over fixed-route.

\subsection{The pros and cons of MOD transit systems}
To further shed light on what shapes respondents' preference for MOD versus fixed-route transit, in this section we discuss the pros and cons of MOD transit systems by examining survey responses to two relevant questions. The survey questions were stated as following: ``Below are some of the benefits/drawbacks associated with the proposed RITMO system, which of them matter to you (please select up to three items)?" Besides a predetermined list of benefits and drawbacks, an ``other, please specify" option was also included to allow open-ended answers. Table 5 presents the responses from all respondents and also responses from the disadvantaged travelers only. Disadvantaged travelers here are defined as individuals who have a household income of less than \$25,000, are 60 years or above, do not own a car, or have a disability. Since the two questions were added to the survey after we launched it in Ypsilanti, only a subset of the Ypsilanti sample answered them. ``Frequency" and ``Percentage" indicates the total number of individuals selecting an item and it as a proportion of the all sample respectively. Overall, the respondents from the disadvantaged travelers are not very different from those from the full sample.
%and so we do not provide separate discussions here.

The results show that the most important benefit of MOD transit perceived by respondents from both samples is the enhanced accessibility it provides, that is, improving access to the number of destinations that individuals can get to using transit. Potential benefits of secondary importance include reductions in walking time, higher flexibility, more comfort (due to being able to wait at home), and service-hour extensions. Notably, a higher proportion of the Ypsilanti respondents valued walking-time reductions and service-hour extensions compared to the Detroit respondents, which is likely because respondents from Ypsilanti Township often live far away from bus stops and receive inadequate transit serviced during early morning, late evening, and the weekends. Finally, a small percentage of respondents indicated that they value the economic efficiency of the RITMO system. Some respondents selected the ``other, please specify" option and suggested some additional benefits of the RITMO system, which includes the ``cool factor" associated with it and the potential to save parking and car ownership and operating costs when drivers decided to switch to transit. 

Regarding the potential drawbacks of the RITMO system, respondents' primary concern seemed to be the process required to use MOD (i.e., actively requesting for a ride). This is likely due to two reasons. First is the travel-behavior inertia effect. A stream of studies have shown that travelers tend to repeat their behavior and are not willing to accept changes unless the new alternative significantly improves their travel experiences \citep{garling2003introduction,chorus2012travel}. The need to actively request for a ride, wait for, and search for the assigned vehicle would be particularly undesirable for many transit riders who are satisfied with the existing fixed-route system and are accustomed to using it. Second is the need to be familiar with modern technology, which is more of a concern for less tech-savvy respondents. Our interviews with some transit professionals working on pilot MOD transit projects in the U.S. revealed that these projects often failed to attract senior users because of the difficulty that they encountered in using ridesharing apps. Although some pilot projects provided the option of requesting for a ride by phone calls, such as the Via/Arlington partnership, the process may be perceived by users as somewhat burdensome. 

Respondents' secondary concerns were potential cost increases and logistic issues such as phone battery running out or lacking internet access or the malfunction of the technology that RITMO depends on. Finally, other issues raised from open-ended responses include uncertainty about service reliability (e.g., unknown wait time, particularly during peak hours), safety concerns (e.g., sitting with strangers in a small vehicle and vehicles going into unsafe neighborhoods), and environmental concerns (e.g., more congestion and green-gas emissions since more small-sized vehicles are required to replace large-volume buses).

\begin{table}[ht!]
\centering
\caption{Important benefits and drawbacks of the proposed RITMO system perceived by respondents}
\footnotesize
\resizebox{1\textwidth}{!}{% <------ Don't forget this %
\begin{tabular}{p{4.5cm}cccccccc}
\hline
                                                                                       & \multicolumn{4}{c}{Detroit data}                                             & \multicolumn{4}{c}{Ypsilanti data}                                           \\ \cline{2-9} 
                                                                                       & \multicolumn{2}{c}{All} & \multicolumn{2}{c}{Disadvantaged} & \multicolumn{2}{c}{All} & \multicolumn{2}{c}{Disadvantaged} \\
                                                                                       
                                                                                       & \multicolumn{2}{c}{sample} & \multicolumn{2}{c}{travelers} & \multicolumn{2}{c}{sample} & \multicolumn{2}{c}{travelers} \\
                                                                                       & \multicolumn{2}{c}{($N=441$)}  & \multicolumn{2}{c}{($N=233$)}               & \multicolumn{2}{c}{($N=251$)}  & \multicolumn{2}{c}{($N=48$)}                \\ \hline
Potential benefits of RITMO that matter to respondents                                      & Freq.     & \%     & Freq.            & \%           & Freq.     & \%     & Freq.            & \%           \\ \hline
It increases the number of places that passengers can get to using transit             & 267           & 60.54\%        & 148                  & 63.52\%              & 161           & 62.89\%        & 25                   & 52.08\%              \\
It reduces the amount of walking (e.g. walking to bus stop) of a transit trip          & 210           & 47.62\%        & 113                  & 48.50\%              & 147           & 57.42\%        & 22                   & 45.83\%              \\
It allows passengers to request a ride whenever they want and wherever they are at     & 220           & 49.89\%        & 96                   & 41.20\%              & 129           & 50.39\%        & 21                   & 43.75\%              \\
It allows passengers to wait home instead of at a bus stop                             & 214           & 48.53\%        & 97                   & 41.63\%              & 128           & 50.00\%        & 17                   & 35.42\%              \\
It can extend transit service hours for early morning/late evening/weekends            & 163           & 36.96\%        & 74                   & 31.76\%              & 137           & 53.52\%        & 9                    & 18.75\%              \\
It can be more economically efficient than a fixed-route bus system                    & 153           & 34.69\%        & 63                   & 27.04\%              & 100           & 39.06\%        & 11                   & 22.92\%              \\
                                                                                       &               &                &                      &                      &               &                &                      &                      \\ \hline
Potential drawbacks of RITMO that matter to respondents                                     & Freq.     & \%     & Freq.            & \%           & Freq.     & \%     & Freq.            & \%           \\ \hline
The cost for a RITMO trip is not likely to be lower than a bus trip                    & 156           & 35.37\%        & 91                   & 39.06\%              & 89            & 35.46\%        & 12                   & 25.00\%              \\
The need to request for a ride instead of just simply waiting for a bus to come        & 200           & 45.35\%        & 104                  & 44.64\%              & 119           & 47.41\%        & 16                   & 33.33\%              \\
Passengers are unable to use it when phone battery runs out or have no internet access & 170           & 38.55\%        & 87                   & 37.34\%              & 78            & 31.08\%        & 19                   & 39.58\%              \\
Potential malfunction of the internet and the RITMO app                                & 146           & 33.11\%        & 68                   & 29.18\%              & 51            & 20.32\%        & 17                   & 35.42\%              \\
Difficulty in finding the ``street corner'' to be picked up                            & 87            & 19.73\%        & 44                   & 18.88\%              & 61            & 24.30\%        & 14                   & 29.17\%              \\ \hline
\end{tabular}
}
\end{table}

\section{Discussion}
\subsection{Policy implications}
The findings of this paper generate insights that can inform transportation policymaking and guide the design of future MOD transit systems to ensure their successful implementation. First, we find that incorporating mobility-on-demand services into the service suite of public transit is likely to gain widespread support among local residents. A majority of our survey respondents favor a proposed integrated MOD transit system over the existing fixed-route system. If the access barriers to MOD services can be eliminated (e.g., through accepting cash fare payment and allowing riders to book trips with phone call or text messages), MOD transit systems can be especially beneficial for the more transit-dependent low-income, aged, carless, and disabled travelers. The integration of MOD with conventional public transit brings the promise of offering affordable and convenient public transit services to areas that were unreachable to disadvantaged travelers. 

%Many of these disadvantaged travelers received inadequate transit services from conventional fixed-route systems, which inhibits them from accessing employment opportunities, healthcare, and health food \citep{lichtenwalter2006examining, walker2010disparities, syed2013traveling} and often forces them to resort to high-cost taxi services to meet necessary travel needs \citep{renne2014socioeconomics}. 
%The new and innovative ride-hailing services operated by private companies are commonly found to provide convenient and reliable services, but disadvantaged travelers are often deterred from using them due to a variety of barriers such as high prices and limited to no access to credit or debit cards, smartphone, internet access, and the skills to use a smartphone app \citep{kodransky2014connecting, dillahunt2017uncovering,dillahunt2018getting}. 

%Identified barriers to participation in shared vehicle services by low-income individuals include a dearth of stations in low-income neighborhoods; transactionally complicated rules of membership and use, requirements to hold credit cards and have Internet access, high prices, lack of information about the new services, and cultural factors, including distrust of authority and discomfort with shared mobility systems \cite{kodransky2014connecting}.

Moreover, we found weaker preference for MOD transit among individuals who had no mobile data plan and among individuals who never heard of or used ride-hailing. If a lack of affordability prevents people from purchasing mobile data, possible solutions to this issue would be installing Wi-Fi access hotspots at key locations or providing subsidies to certain populations for mobile data plan purchases. More likely, however, we believe the underlying factor that causes these individuals to have a less favorable view on MOD to be the same, which is a lack of technology proficiency; that is, less tech-savvy individuals do not use the internet with mobile devices frequently and also tend to have no experience with the emerging ride-hailing services. It is natural for less technology-savvy individuals to express less support for MOD transit due to their perceived difficulty of use associated with on-demand ridesharing services. Consequently, addressing this digital divide and the unwillingness of some transit riders to adopt new technologies should be a priority for transit operators. Possible measures include targeting marketing and outreach programs to less technology-proficient individuals, making easy-to-understand materials to guide the use of MOD transit serivices, and incorporating more user-friendly features (e.g., voice activation) into the MOD smartphone app, etc.

%our results show that the main access barrier to MOD adoption appears to be technology illiteracy instead of lacking infrastructure access (e.g., lacking access to bank accounts and smartphones). This is because compared to other population groups, weaker preference for MOD transit is found among individuals who never heard of or never used ride-hailing and individuals who had no data plan on their phone, but not among individuals who had no bank accounts or smartphones. When many individuals decided not to purchase a data plan, it is often because they do not use the internet and are not proficient with modern technology. Therefore, we suggest transit agencies to devote considerable efforts to accommodate the needs of less tech-savvy individuals. Possible measures include incorporating more user-friendly features (e.g., voice activation) into the MOD smartphone app, 

Also, females have more reservations for MOD transit than males. We obtained scant evidence from the survey that suggests that such preference difference arose from women's safety concerns, e.g., being uncomfortable with sitting with strangers in a small-size vehicle and fearing that on-demand shuttles travelling to unsafe places. Transit agencies should thus address such concerns with corresponding measures, such as putting larger space gaps between seats, installing security cameras, and ensuring adequate driver training. In addition, the MOD transit concept appears to draw similar level of support across different age, income level, and racial groups, assuming their level of technology proficiency and the current transit services they receive are similar. This findings suggests that while young to middle-aged and college-educated individuals are more likely to the first adopters of MOD transit services, other population groups would follow suit if the digital divide gap is narrowed and if MOD significantly improves transit experiences.

Finally, pilot MOD initiatives should first target areas that are not receiving adequate transit services from the existing fixed-route system. As expected, we found that respondents living at places well-served by the fixed-route services are less supportive of the MOD transit concept than those who are not receiving adequate services. This further confirms that the biggest potential of MOD transit is providing services to previously under-served areas (i.e., low density areas or places lacking convenient last-mile transit access), which was also the most important benefit of MOD transit perceived by our survey respondents. To identify transit-deficit areas to test MOD pilots, a transit operator may apply several criteria such as being distant from transit stops and having low job accessibility by transit (number of jobs reachable within a certain amount of transit travel time).

%Finally, the transition from a conventional fixed-route system to a MOD transit system is best implemented incrementally. Although our results show overwhelming community support for the concept of integrating MOD services with public transit, a significant proportion of the respondents still favor the conventional fixed-route system over an integrated MOD transit system. Oppositions not only arise from a digital divide concern, which was already discussed above, but also travelers' behavior inertia. Many transit riders who are accustomed to using the fixed-route services would be unwilling to change their current travel patterns unless there are significant improvements to the quality of transit services. The need to actively request a ride is a particularly undesirable feature of MOD transit, particular for individuals who are less proficient with technology. Transit agencies should thus implement incremental changes at first and let transit riders gradually adapt to a new form of transit services.

\subsection{Limitations}
A major limitation of this study is that our web-based survey is likely to impede participation from individuals who are uncomfortable completing online surveys  (e.g., due to limited computer self-efficacy or perceived difficulty in using digital devices) or lack access to the internet and/or digital devices in the first place. As studies on the sharing economy have shown, these individuals tend to be less willing to adopt services of the sharing economy such as real-time ride sharing \citep{hsiao2018role}. Their under-representation in the survey sample may thus cause an overall perception of MOD transit to have a positive bias. Our on-site recruitment efforts in Detroit mitigated this problem but did not solve it. 

In addition, we assumed that future MOD transit systems would allow passengers to request rides by phone calls as an alternative to booking through a smartphone app or internet web page. This could explain why we found that individuals without a smartphone, a mobile data plan, or internet access at home did not have weaker preferences for MOD transit compared to individuals facing no such constraints, because most of the former individuals reported owning a cellphone. In reality, when a transit agency deploys a MOD transit system, the phone-call option may be unavailable or less convenient to use. On the positive side, however, our finding can also be interpreted as suggesting that adding a call-in option could effectively mitigate the technology-access problem for adopting MOD transit.

Another limitation is the hypothetical nature of the proposed MOD transit system. Like any stated-preference data, the results should be interpreted with caution, as individuals' actual behavior may be different from their responses under hypothetical situations \citep{diamond1994contingent}. Particularly, for individuals with limited experience in using MOD services and thus do not have a full picture of its benefits and drawbacks, their stated responses at present can lack validity; their preferences for MOD transit in the future can be very different after fully experiencing the proposed MOD transit system. Notably, there is also a status quo bias in respondents' preferences. Their perception of the proposed RITMO system is likely shaped by their perception of the existing ride-hailing and public transit services. If they hold a negative perception of the existing transit services, the expressed preference for MOD transit over fixed-route may simply be a desire for change. Also, the positive perception of the existing privately-operated ride-hailing services in terms of its service quality and reliability may lead to a positive bias toward the MOD transit concept, because MOD transit is not likely to offer the same level of service compared to for-profit private ride-hailing services. 

Moreover, our survey elicits individual responses to a blueprint of the integrated MOD plus conventional transit system, which is something distant to them and so less likely to result in strong oppositions. Urban planners often find that local communities would support a master plan but raise oppositions during the on-the-ground implementation stage when their imminent interests are on the line. This suggests the need for future studies to examine existing MOD transit pilot projects, the results of which should shed further light on the obstacles that prevent the switch from conventional public transit systems to the MOD transit systems.

\section{Conclusion}

The rapid rise of ride-hailing (e.g., Uber and Lyft) and microtransit (e.g., Via and Chariot) services and the development of autonomous vehicles lead many to speculate their implications on the future of public transit. The speculated possible future scenarios range from the complete demise of public transit as cheap and convenient autonomous-vehicle trips make transit obsolete, to a segmented market where fixed-route transit serves the highest-demand corridors while autonomous vehicles occupy other markets, to a resurgence of public transit as new technology renders private auto ownership unnecessary \citep{polzin2016implications}. Faced with many uncertainties, public transit needs to develop a vision for its future and look for creative ways to improve the service quality and operation efficiency in order to stay competitive. Many transit observers have suggested that an ideal future transit system should integrate conventional fixed-route with mobility-on-demand, with the former serving busy corridors, the latter serving lower-density areas and providing last-mile feeder service to transit stops, and the operation of the two being synchronized by a central control system.
%An ideally integrated MOD transit system has the potential to maximize the economic efficiency of the transport network and to provide convenient and affordable services to the general public with improved service levels and wider geographic coverage.

Realizing such an integrated MOD transit system requires forward thinking and active planning from transit operators, and they should carefully consider a range of issues such as network design, organizational integration, operating model, and user buy-in, etc. To complement previous work that has primarily focused on the operation and system design aspects of MOD transit, this paper generates insights on user perceptions and preferences. We conducted a survey in two low-source communities, namely Detroit and Ypsilanti, Michigan, to investigate how residents, particularly the more transit-dependent, disadvantaged travelers, and the less technology-savvy ones, would react to the concept of an integrated MOD transit system. 
%e  in which survey participants were asked to indicate their preference for a proposed MOD transit system (named RITMO transit) versus the existing fixed-route transit system on a five-response-category rank order scale. Response categories include I stronger prefer RITMO over fixed-route, I sort of prefer RITMO over fixed-route, not so sure, I sort of prefer fixed-route over RITMO, and I strongly prefer fixed-route over RIMTO.

A majority of the survey respondents indicated that they strongly or sort of prefer the RITMO system over the existing fixed-route system, whereas a small minority would rather stick with the conventional fixed-route transit system. We also found that males, college graduates, individuals who have never heard of or used ride-hailing before, and individuals who currently receive inferior transit service from the fixed-route system are more likely to hold a more favorable perception of the RITMO system. Moreover, preferences appeared to vary little by age, income, race, or disability status. To our surprise, lacking access to a bank account, a smartphone, or internet at home or having a disability were not associated with individual preference for MOD transit versus fixed-route, but lacking access to a mobile data plan was negatively associated with it. Unsurprisingly, the survey respondents view the most important benefit of MOD transit systems as the enhanced transit accessibility, i.e., the potential to reach more places with public transit. Their major concerns, ranked in order, include the need to actively request for rides (as opposite to learn the schedule and wait for buses to come themselves), possible transit-fare increases, and concerns regarding technological failures. 

%These results are statistical significant in both the Detroit-data model and the Ypsilanti-data model. 
%Moreover, older and car-owning respondents appeared to be less supportive a switch from fixed-route transit system to a MOD one. 
%A major limitation of our study is that the data collected from respondents are based on their responses to a hypothetical MOD transit system, which lack the validity resided in data based on individuals' real-life experiences and actual behavior. In other words, individuals' overall perception of a MOD transit system and the perceived benefits and drawbacks associated with it may be different if they had really experienced it. Therefore, 
Future research that examine the existing MOD transit initiatives would be helpful to verify the results of this study and to generate further insights on the pros and cons of MOD transit systems. Future studies should also evaluate and quantify the potential of MOD transit systems to enhance transit accessibility to essential destinations, lower car use, and reduce parking demand. Finally, there is a need for future research to identify effective solutions to overcome the barriers to adopt MOD transit and to explore user-friendly design features that can lower the technological proficiency requirement in using it.

\begin{acknowledgements}
This project is funded by Poverty Solutions and Michigan Institute for Data Science at the University of Michigan and Grant 7F-30154 from the U.S. Department of Energy. The authors would thank Jonathan Levine for providing helpful comments and Ali Shahin for his research assistance.
\end{acknowledgements}

\newpage
% BibTeX users please use one of
\bibliographystyle{spbasic}      % basic style, author-year citations
%\bibliographystyle{spmpsci}      % mathematics and physical sciences
%\bibliographystyle{spphys}       % APS-like style for physics
%\bibliography{}   % name your BibTeX data base

\bibliography{sample}

\begin{thebibliography}{34}
\providecommand{\natexlab}[1]{#1}
\providecommand{\url}[1]{{#1}}
\providecommand{\urlprefix}{URL }
\expandafter\ifx\csname urlstyle\endcsname\relax
  \providecommand{\doi}[1]{DOI~\discretionary{}{}{}#1}\else
  \providecommand{\doi}{DOI~\discretionary{}{}{}\begingroup
  \urlstyle{rm}\Url}\fi
\providecommand{\eprint}[2][]{\url{#2}}

\bibitem[{Alemi et~al(2018)Alemi, Circella, Handy, and
  Mokhtarian}]{alemi2018uber}
Alemi F, Circella G, Handy S, Mokhtarian P (2018) What influences travelers to
  use uber? exploring the factors affecting the adoption of on-demand ride
  services in california. Travel Behaviour and Society 13:88--104

\bibitem[{Atasoy et~al(2015)Atasoy, Ikeda, Song, and
  Ben-Akiva}]{atasoy2015concept}
Atasoy B, Ikeda T, Song X, Ben-Akiva ME (2015) The concept and impact analysis
  of a flexible mobility on demand system. Transportation Research Part C:
  Emerging Technologies 56:373--392

\bibitem[{Buehler(2018)}]{buehler2018av}
Buehler R (2018) Can public transportation compete with automated and connected
  cars? Journal of Public Transportation 21(1):2

\bibitem[{Chorus and Dellaert(2012)}]{chorus2012travel}
Chorus CG, Dellaert BG (2012) Travel choice inertia: the joint role of risk
  aversion and learning. Journal of Transport Economics and Policy (JTEP)
  46(1):139--155

\bibitem[{Clewlow and Mishra(2017)}]{clewlow2017ridehailing}
Clewlow RR, Mishra GS (2017) Disruptive transportation: The adoption,
  utilization, and impacts of ride-hailing in the united states. University of
  California, Davis, Institute of Transportation Studies, Davis, CA, Research
  Report UCD-ITS-RR-17-07

\bibitem[{Diamond and Hausman(1994)}]{diamond1994contingent}
Diamond PA, Hausman JA (1994) Contingent valuation: is some number better than
  no number? Journal of economic perspectives 8(4):45--64

\bibitem[{Dias et~al(2017)Dias, Lavieri, Garikapati, Astroza, Pendyala, and
  Bhat}]{dias2017behavioral}
Dias FF, Lavieri PS, Garikapati VM, Astroza S, Pendyala RM, Bhat CR (2017) A
  behavioral choice model of the use of car-sharing and ride-sourcing services.
  Transportation 44(6):1307--1323

\bibitem[{Dillahunt and Veinot(2018)}]{dillahunt2018getting}
Dillahunt TR, Veinot TC (2018) Getting there: Barriers and facilitators to
  transportation access in underserved communities. ACM Transactions on
  Computer-Human Interaction (TOCHI) 25(5):29

\bibitem[{Dillahunt et~al(2017)Dillahunt, Kameswaran, Li, and
  Rosenblat}]{dillahunt2017uncovering}
Dillahunt TR, Kameswaran V, Li L, Rosenblat T (2017) Uncovering the values and
  constraints of real-time ridesharing for low-resource populations. In:
  Proceedings of the 2017 CHI Conference on Human Factors in Computing Systems,
  ACM, pp 2757--2769

\bibitem[{Durand et~al(2018)Durand, Harms, Hoogendoorn-Lanser, and
  Zijlstra}]{durand2018mobility}
Durand A, Harms L, Hoogendoorn-Lanser S, Zijlstra T (2018)
  Mobility-as-a-service and changes in travel preferences and travel behaviour:
  a literature review

\bibitem[{Errico et~al(2013)Errico, Crainic, Malucelli, and
  Nonato}]{errico2013survey}
Errico F, Crainic TG, Malucelli F, Nonato M (2013) A survey on planning
  semi-flexible transit systems: methodological issues and a unifying
  framework. Transportation Research Part C: Emerging Technologies 36:324--338

\bibitem[{Feigon and Murphy(2016)}]{feigon2016shared}
Feigon S, Murphy C (2016) Shared mobility and the transformation of public
  transit. Transportation Research Board, Transit Cooperative Research Program
  Research Report 188

\bibitem[{G{\"a}rling and Axhausen(2003)}]{garling2003introduction}
G{\"a}rling T, Axhausen KW (2003) Introduction: Habitual travel choice.
  Transportation 30(1):1--11

\bibitem[{Ge et~al(2016)Ge, Knittel, MacKenzie, and Zoepf}]{ge2016racial}
Ge Y, Knittel CR, MacKenzie D, Zoepf S (2016) Racial and gender discrimination
  in transportation network companies. Tech. rep., National Bureau of Economic
  Research

\bibitem[{Hensher(2017)}]{hensher2017future}
Hensher DA (2017) Future bus transport contracts under a mobility as a service
  (maas) regime in the digital age: Are they likely to change? Transportation
  Research Part A: Policy and Practice 98:86--96

\bibitem[{Hsiao et~al(2018)Hsiao, Moser, Schoenebeck, and
  Dillahunt}]{hsiao2018role}
Hsiao JCY, Moser C, Schoenebeck S, Dillahunt TR (2018) The role of
  demographics, trust, computer self-efficacy, and ease of use in the sharing
  economy. In: Proceedings of the 1st ACM SIGCAS Conference on Computing and
  Sustainable Societies, ACM, p~37

\bibitem[{Kodransky and Lewenstein(2014)}]{kodransky2014connecting}
Kodransky M, Lewenstein G (2014) Connecting low-income people to opportunity
  with shared mobility. Report produced for Living Cities New York, NY:
  Institute for Transportation \& Development Policy

\bibitem[{Mah{\'e}o et~al(2017)Mah{\'e}o, Kilby, and
  Van~Hentenryck}]{maheo2017benders}
Mah{\'e}o A, Kilby P, Van~Hentenryck P (2017) Benders decomposition for the
  design of a hub and shuttle public transit system. Transportation Science

\bibitem[{Mulley and Kronsell(2018)}]{mulley2018workshop}
Mulley C, Kronsell A (2018) Workshop 7 report: The “uberisation” of public
  transport and mobility as a service (maas): Implications for future
  mainstream public transport. Research in Transportation Economics 69:568--572

\bibitem[{{{Pew Research Center}}(2018)}]{Pew2018}
{{Pew Research Center}} (2018) Mobile fact sheet. Pew Research Center Survey
  Results Accessed on December 29, 2018 from
  http://wwwpewinternetorg/fact-sheet/mobile/

\bibitem[{Polzin(2016)}]{polzin2016implications}
Polzin SE (2016) Implications to public transportation of emerging
  technologies. National Center for Transit Research, University of South
  Florida

\bibitem[{Schwieterman et~al(2018)Schwieterman, Livingston, and Van
  Der~Slot}]{schwieterman}
Schwieterman JP, Livingston M, Van Der~Slot S (2018) Partners in transit: A
  review of partnerships between transportation network companies and public
  agencies in the united states. Tech. rep., Chaddick Institute for
  Metropolitan Development, Depaul University, Policy Series

\bibitem[{Shaheen and Chan(2016)}]{shaheen2016lastmile}
Shaheen S, Chan N (2016) Mobility and the sharing economy: Potential to
  facilitate the first-and last-mile public transit connections. Built
  Environment 42(4):573--588

\bibitem[{Shaheen et~al(2017)Shaheen, Bell, Cohen, and
  Yelchuru}]{shaheen2017equity}
Shaheen S, Bell C, Cohen A, Yelchuru B (2017) Travel behavior: Shared mobility
  and transportation equity. Tech. rep., Federal Highway Administration,
  Technical Report PL-18-007

\bibitem[{Shen et~al(2018)Shen, Zhang, and Zhao}]{shen2018integrating}
Shen Y, Zhang H, Zhao J (2018) Integrating shared autonomous vehicle in public
  transportation system: A supply-side simulation of the first-mile service in
  singapore. Transportation Research Part A: Policy and Practice 113:125--136

\bibitem[{Smith(2016)}]{smith2016shared}
Smith A (2016) Shared, collaborative and on demand: The new digital economy.
  Pew Research Center

\bibitem[{Sochor et~al(2015)Sochor, Str{\"o}mberg, and
  Karlsson}]{sochor2015users}
Sochor JL, Str{\"o}mberg H, Karlsson M (2015) Challenges in integrating user,
  commercial, and societal perspectives in an innovative mobility service. In:
  Proceedings of the 94th Annual Meeting of the Transportation Research Board,
  Washington, DC January 11-15, 2015

\bibitem[{Stiglic et~al(2018)Stiglic, Agatz, Savelsbergh, and
  Gradisar}]{stiglic2018enhancing}
Stiglic M, Agatz N, Savelsbergh M, Gradisar M (2018) Enhancing urban mobility:
  Integrating ride-sharing and public transit. Computers \& Operations Research
  90:12--21

\bibitem[{Str{\"o}mberg(1996)}]{stromberg1996collapsing}
Str{\"o}mberg U (1996) Collapsing ordered outcome categories: a note of
  concern. American journal of epidemiology 144(4):421--424

\bibitem[{Tussyadiah and Pesonen(2018)}]{tussyadiah2018drivers}
Tussyadiah IP, Pesonen J (2018) Drivers and barriers of peer-to-peer
  accommodation stay--an exploratory study with american and finnish
  travellers. Current Issues in Tourism 21(6):703--720

\bibitem[{Walker(2012)}]{walker2012human}
Walker J (2012) Human transit: How clearer thinking about public transit can
  enrich our communities and our lives. Island Press

\bibitem[{Wang et~al(2014)Wang, Quddus, Enoch, Ryley, and
  Davison}]{wang2014DRT}
Wang C, Quddus M, Enoch M, Ryley T, Davison L (2014) Multilevel modelling of
  demand responsive transport (drt) trips in greater manchester based on
  area-wide socio-economic data. Transportation 41(3):589--610

\bibitem[{Williams(2016)}]{williams2016generalized}
Williams R (2016) Understanding and interpreting generalized ordered logit
  models. The Journal of Mathematical Sociology 40(1):7--20

\bibitem[{Yan et~al(2018)Yan, Levine, and Zhao}]{yan2018integrating}
Yan X, Levine J, Zhao X (2018) Integrating ridesourcing services with public
  transit: An evaluation of traveler responses combining revealed and stated
  preference data. Transportation Research Part C: Emerging Technologies
  \doi{https://doi.org/10.1016/j.trc.2018.07.029},
  \urlprefix\url{http://www.sciencedirect.com/science/article/pii/S0968090X18310398}

\end{thebibliography}

% Non-BibTeX users please use
%\begin{thebibliography}{}
%
% and use \bibitem to create references. Consult the Instructions
% for authors for reference list style.
%
%\bibitem{RefJ}
% Format for Journal Reference
%Author, Article title, Journal, Volume, page numbers (year)
% Format for books
%\bibitem{RefB}
%Author, Book title, page numbers. Publisher, place (year)
% etc
%\end{thebibliography}

\end{document}